\documentclass[showpacs,preprint,epsfig,floats,aps]{revtex4}
\usepackage{epsfig}

\begin{document}

\title{$e^-e^+\rightarrow\overline NN$ at Threshold and Proton Form Factor}
\author{Y. Yan \footnotemark[1]\footnotemark[2],
  K. Khosonthongkee \footnotemark[1]\footnotemark[2],
  C. Kobdaj \footnotemark[1]\footnotemark[2],
  P. Suebka \footnotemark[1]\footnotemark[2]}
\affiliation{ \footnotemark[1] School of Physics,
Suranaree University of Technology, Nakhon Ratchasima
30000, Thailand
\\
\footnotemark[2] Thailand Center of Excellence in Physics, Ministry
of Education, Bangkok, Thailand \email{yupeng@sut.ac.th}
\vspace*{0.4\baselineskip}}

\begin{abstract}
The reactions $e^+e^-\rightarrow \overline NN$ at energy threshold
are studied in a non-perturbative quark model. The puzzling experimental result that
$\sigma(e^+e^-\rightarrow\overline
pp)/\sigma(e^+e^-\rightarrow\overline nn) < 1$ can be understood
in the framework of the phenomenological nonrelativistic quark
model and the theoretical
predictions for the time-like proton form factor at energy threshold
are well consistent with the experimental data. The work suggests
that the two-step process, in which the primary $\overline qq$ pair
forms first a vector meson which in turn decays into a hadron pair,
is dominant over the one-step process in which the primary
$\overline qq$ pair is directly dressed by additional $\overline qq$
pairs to form a hadron pair.
The experimental data on the reactions
$e^+e^-\rightarrow\overline nn$ and $\overline pp$ strongly suggest
the reported vector meson $\omega(1930)$ to be a $2D$-wave particle, while the
vector meson $\rho(2000)$ is preferred to be a mixture of $3S$ and
$2D$ states.
\end{abstract}

\maketitle

\section{Introduction}
\begin{figure}[h]
 \centering
 \includegraphics[width=0.7\textwidth]{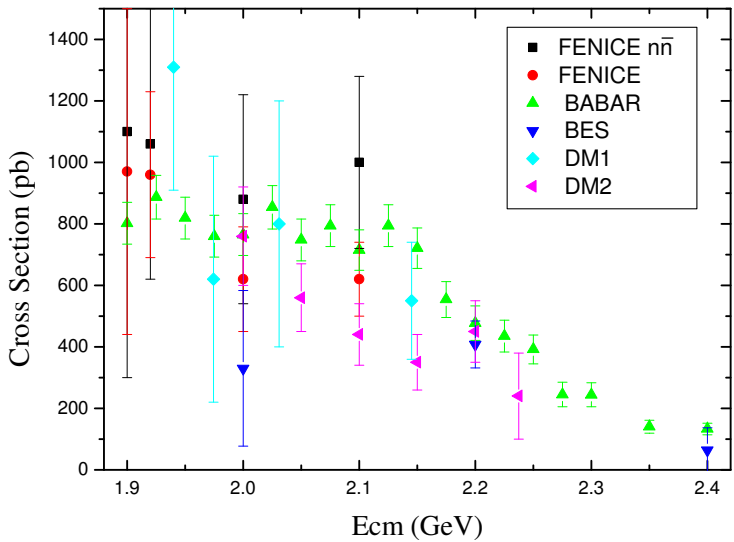}
\caption{Comparison of the cross sections for
$e^+e^-\rightarrow\overline pp$ and $e^+e^-\rightarrow\overline nn$
at the $\overline NN$ threshold region. The
squares~\cite{FENICEnn} are for the reaction $e^+e^-\rightarrow\overline nn$ while other
symbols are for the
reaction $e^+e^-\rightarrow\overline pp$.
\label{fig1}}
\end{figure}

Experimental data on the reaction $e^+e^-\rightarrow\overline nn$
from the FENICE collaboration \cite{FENICEnn}, earlier data on the
reaction $e^+e^-\rightarrow\overline pp$ from the FENICE \cite{FENICEpp} and DM2
collaborations \cite{DM2}, data collected at the LEAR antiproton
ring at CERN on the time-reversed reaction $\overline pp\rightarrow
e^+e^-$ \cite{LEAR}, and also recent measurements from the BES
Collaboration \cite{BES} and the BaBar Collaboration \cite{BaBar}, which are
summarized in FIG.\ref{fig1}, indicate at energies around the
$\overline NN$ threshold with $E_{\rm c.m.}\sim 2$ GeV
\begin{eqnarray}\label{eq::1}
\frac{\sigma(e^+e^-\rightarrow\overline
pp)}{\sigma(e^+e^-\rightarrow\overline nn)} <1.
\end{eqnarray}
Averaging over the FENCICE, DM2 and LEAR data \cite
{FENICEnn,FENICEpp,DM2,LEAR} on both the direct and time-reversed
reactions, the work \cite{Ellis} finds that the ratio is
$0.66^{+0.16}_{-0.11}$.

In a naive perturbative description of electron-positron
annihilation into baryons the virtual time-like photon first decays
into a quark-antiquark pair, then the quark-antiquark pair is
dressed by two additional quark-antiquark pairs excited out of the
vacuum to form a baryon pair. The dressing process does not
distinguish between u and d quarks at high momentum transfers since
in the description of perturbative QCD the gluon couplings are
flavor blind. In the conventional perturbative picture the only
difference between the proton and neutron production arises from the
different electric charges of the primary quark-antiquark pairs. One
expects to get
\begin{eqnarray}\label{eq::2}
\frac{\sigma(e^+e^-\rightarrow\overline
pp)}{\sigma(e^+e^-\rightarrow\overline nn)} > 1
\end{eqnarray}
at large momentum transfers where the $u$ quark contribution
dominates in the proton and the $d$ quark in the neutron.

The reaction $e^+e^-\rightarrow\overline NN$ at energies around the
$\overline NN$ threshold is highly nonperturbative, hence the
problem must be tackled in a nonperturbative manner. In this work we
model the reactions by the nonperturbative $^3P_0$ quark dynamics
which describes quark-antiquark annihilation and creation. It was
shown that the $^3P_0$ approach is phenomenologically successful in
the description of hadronic
couplings~\cite{Yaouanc,Tueb,barnes}.

The reaction $e^+e^-\rightarrow\overline NN$ may arise from two
different processes: (1) the primary $\overline qq$ pair is dressed
directly by two additional quark-antiquark pairs created out of the
vacuum to form a baryon pair; and (2) the primary $\overline qq$
pair forms a virtual vector meson first, then the virtual vector
meson decays into a baryon pair. We expect that the second process
is dominant over the first because of the considerable success of
the vector dominance model. The importance of the second process is
also discussed in the work \cite{Ellis}. Our work is arranged as
follows: We study the reaction $e^+e^-\rightarrow\overline NN$ in
the one-step and two-step processes in Sec. II and III,
respectively. We give our conclusions in Sec. IV.

\section{Reaction $e^+e^- \rightarrow \overline NN$ in one-step process}
\begin{figure}[h]
\centering
\includegraphics[height=0.2\textheight]{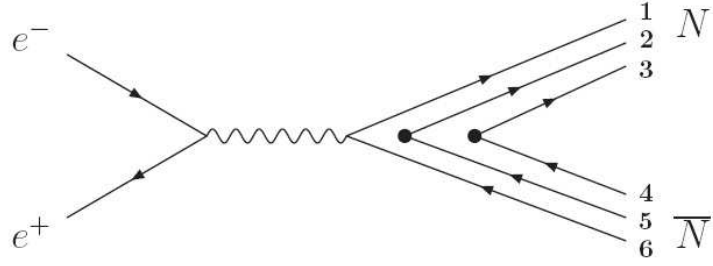}
\caption{One-step process in which the primary $\overline qq$ pair
is directly dressed by additional $\overline qq$ pairs to form a
hadron pair.\label{fig2}}
\end{figure}

The reaction $e^+e^-\rightarrow\overline NN$ may arise from the
following process:
the $e^+e^-$ pair
annihilates into a virtual time-like photon, the virtual photon
decays into a $\overline qq$ pair, and finally the $\overline qq$
pair is dressed by two additional quark-antiquark pairs created out of
the vacuum to form a nucleon-antinucleon pair, as shown in FIG.\ref{fig2}.
We refer to this process as the one-step
reaction.
The transition amplitude is expressed formally as
\begin{eqnarray}\label{eq::3}
T=\langle\overline NN|V_{\overline qq}|\overline qq\rangle
\langle\overline qq|G|\overline qq\rangle\langle\overline
qq|T|e^+e^-\rangle
\end{eqnarray}
where $\langle\overline qq|T|e^+e^-\rangle$ is simply the
transition amplitude of $e^+e^-$ to a primary quark pair,
$\langle\overline qq|G|\overline qq\rangle$ is the Green function
describing the propagation of the intermediate $\overline qq$
state and $\langle\overline NN|V_{\overline qq}|\overline qq\rangle$
denotes the amplitude of the process of a $\overline qq$ pair to a
$\overline NN$ pair. $V_{\overline qq}$ is the effective vertex for
creation and destruction of two quark-antiquark pairs in quark
models, which is identified in the context of the $^3P_0$
quark-antiquark dynamics. The effective vertex $V_{\overline qq}$ takes
the form
\begin{eqnarray}\label{eq::4}
V_{\overline qq} \equiv V_{25}V_{34}
\end{eqnarray}
 with
\begin{eqnarray} \label{eq::5}
    V_{ij}&=&\lambda_{ij}\vec\sigma_{ij}\cdot(\vec p_i-\vec
    p_j)\hat{F}_{ij}\hat{C}_{ij}\delta(\vec p_i+\vec
    p_j)\nonumber\\
    &=&\lambda_{ij}\sum_\mu\sqrt\frac{4\pi}{3}(-1)^\mu\sigma_{ij}^{-\mu} Y_{1\mu}
    (\vec p_i-\vec p_j)\hat F_{ij}\hat C_{ij}\delta(\vec p_i+\vec
    p_j)
\end{eqnarray}
where $\lambda_{ij}$ is the effective coupling constant for creating quark $i$
and antiquark $j$, $\vec p_i$ and $\vec p_j$
are the momenta of quark and antiquark
created out of the vacuum.  $\sigma_{ij}^{-\mu}$, $\hat F_{ij}$, $\hat C_{ij}$
are respectively
the the spin, flavor and color operators, projecting a
quark-antiquark pair to the respective vacuum quantum numbers.
The operations of flavor, color, and
spin operators onto a $q\overline q$ pair are
\begin{eqnarray}\label{eq::6}
\langle 0,0|\hat{F}_{ij}|\left[\,\overline t_i\otimes
t_j\right]_{T,T_z}\rangle &=&
\sqrt 2\delta_{T,0}\delta_{T_z,0}, \nonumber \\
\langle 0,0|\hat{C}_{ij}|q_\alpha^i\bar q_\beta^j\rangle &=&
    \delta_{\alpha\beta}, \nonumber \\
\langle
0,0|\sigma_{ij}^{-\mu}|\left[\,\overline\chi_i\otimes\chi_j\right]_{JM}\rangle
&=&
    (-1)^M\sqrt 2\delta_{J,1}\delta_{M,\mu}
\end{eqnarray}
where $t_i (\overline t_i)$ and $\chi_i (\overline\chi_i)$ are respectively
the flavor and
spin states of quark (antiquark),
$\alpha$ and $\beta$ the color indices,
and $T$ and $T_z$ the total isospin and its projection of the $q\overline q$ pair.

To evaluate the transition amplitude one needs also a knowledge
of the radial wave function of nucleons. In our calculation, we
simply let nucleons take a radial wave function of the Gaussian
form.
At the $\overline NN$ threshold the $^3P_0$ model gives
\begin{eqnarray}\label{eq::7}
\frac{\sigma(e^+e^-\rightarrow\overline
pp)}{\sigma(e^+e^-\rightarrow\overline nn)} =\frac{81}{16}
\end{eqnarray}
This is a very general result, independent of the effective coupling constants
and also independent of the size parameter of the Gaussian-type nucleon radial
wave function.
The result is clearly not in line with the experimental data, and hence one may
conclude that the one-step process can NOT be dominant for $e^+e^- \rightarrow
\overline N N$ at threshold in the $^3P_0$ model.
\section{Reaction $e^+e^- \rightarrow \overline NN$ in two-step process}
\begin{figure}[h]
\centering
\includegraphics[height=0.15\textheight]{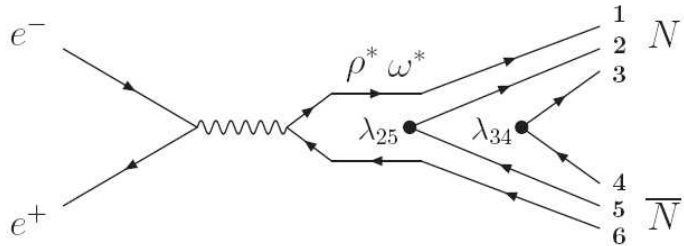}
\caption{Two-step process in which the primary $\overline qq$ pair
forms a virtual vector meson first, then the virtual vector
meson decays into a baryon pair.\label{fig3}}
\end{figure}

In addition to the one-step process, the reaction $e^+e^-\rightarrow\overline NN$ may
arise from the process: the primary $\overline qq$
pair forms a virtual vector meson first, then the virtual vector
meson decays into a baryon pair (as shown in FIG.\ref{fig3}),
as in the context of the vector dominance
model.
We refer to this process as the two-step
reaction.
The corresponding transition amplitude then takes
the form
\begin{eqnarray}\label{eq::8}
T=\langle\pi^+\pi^-| V_{\overline qq}
|\rho\rangle\langle\rho^*(\omega^*)|G|\rho(\omega^*)\rangle\langle\rho^*(\omega^*)
|\overline
qq\rangle\langle\overline qq|T|e^+e^-\rangle
\end{eqnarray}
where $\langle\overline qq|T|e^+e^-\rangle$ is the
transition amplitude of $e^+e^-$ to a primary quark pair,
$\langle\rho^*(\omega^*)|\overline qq\rangle$ is simply the wave
function of the intermediate vector meson $\rho^*(\omega^*)$,
$\langle\rho^*(\omega^*)|G|\rho^*(\omega^*)\rangle$ the Green function describing the
propagation of the intermediate meson $\rho^*(\omega^*)$, and
$\langle\overline NN|V_{\overline qq}|\rho^*(\omega^*)\rangle$ the transition
amplitude of $\rho^*(\omega^*)$ annihilation into a $\overline NN$ pair.
$V_{\overline qq}$ is the effective vertex for
creation and destruction of two quark-antiquark pairs in the $^3P_0$ quark
models, taking
the form
\begin{eqnarray}\label{eq::9}
V_{\overline qq} \equiv V_{25}\;\frac{1}{\Delta E}\;V_{34}
\end{eqnarray}
where $1/\Delta E$, accounting for the energy propagation between
the two $q\overline q$ vertices, is simply set by associating an equal share
of $E_{\rm c.m.}$ to each valence quark \cite{thomas,yankaon,yanNNbar}, and $V_{ij}$ as defined in eq. (\ref{eq::5}).

Based on the evaluations for the reactions
$\rho^*(\omega^*)\rightarrow\overline NN$ and $\rho^*(\omega^*)\rightarrow e^+e^-$, it
is straightforward to work out the transition amplitude of the two-step
diagram shown in FIG.\ref{fig3} for the reaction
$e^+e^- \rightarrow \overline NN$ with the intermediate mesons $\rho^*$ or
$\omega^*$
\begin{eqnarray}
T_{e^+e^- \rightarrow \overline NN}&=&
T_{\rho^*(\omega^*)\rightarrow\overline NN}\frac{1}{E-M_{\rho^*(\omega^*)}}
T_{e^+e^-\rightarrow\rho^*(\omega^*)}.
\end{eqnarray}
The transition amplitude for the process
$\rho^*(\omega^*)\rightarrow e^+e^-$ is
\begin{eqnarray}\label{rhotoee1}
T_{\rho^*(\omega^*)\rightarrow e^+e^-} &=&  \langle
e^+e^-|T|q\overline q\rangle\langle q\overline
q|\rho^*(\omega^*)\rangle \nonumber
\\
&=&\int \frac{d\vec p_q\, d\vec p_{\overline q}}{(2\pi)^{3/2}2E_q}\,
\delta(\vec p_q+\vec p_{\overline q})\, \psi_{\rho^*(\omega^*)}(\vec
p_q,\vec p_{\overline q})\,T_{q\overline q\rightarrow e^+e^-}(\vec
p_q,\vec p_{\overline q})
\end{eqnarray}
where $\psi_{\rho^*(\omega^*)}$ is the wave function of the $\rho^* (\omega^*)$
meson in momentum
space.  The delta function $\delta(\vec p_q+\vec
p_{\overline q})$ indicates that we work in the $\rho^*(\omega^*)$ meson rest
frame. $T_{q\bar q\to e^+e^-}\equiv\langle
e^+e^-|T|q\bar q\rangle$ is the transition amplitude of the reaction
of a quark-antiquark pair to an electron-positron pair, taking the
form
\begin{equation}\label{eq::6}
\langle e^+e^-|T|q\bar q\rangle=-\frac{e_qe}{s}\bar
    u_e(p_{e^-},m_{e^-})\gamma^\mu v_e(p_{e^+},m_{e^+})
    \bar v_q(p_{\bar q},m_{\bar q})\gamma_\mu
    u_q(p_{q},m_{q})
\end{equation}
where $s=(p_q+p_{\bar q})^2$, $e_q$ is the quark charge, and the
Dirac spinors are normalized according to $\bar uu=\bar vv=2m_q$.
Note that only the $P$-wave contributes to the process $e^+e^- \to
\overline NN$ since the spin of the intermediate $\rho^*(\omega^*)$ is 1.

\begin{table}
\begin{center}
\caption{Vector mesons with masses up to the $\overline NN$ threshold.}
\vskip 0.5cm
\label{table1}
\begin{tabular}{|c|c|c|}
\hline
1S & $\rho(770)$ & $\omega(782)$  \\
\hline
2S & $\rho(1450)$ & $\omega(1420)$ \\
\hline
1D & $\rho(1700)$ & $\omega(1650)$   \\
\hline
3S or 2D & $\rho(2000)$ & $\omega(1930)$ \\
\hline
\end{tabular}
\end{center}
\end{table}

Listed in Table \ref{table1} are confirmed and reported vector mesons up to the $\overline NN$
threshold \cite{particletable,rho2000,omega1930}. Among them
$\rho(2000)$ and $\omega(1930)$ are close to the $\overline NN$ threshold and hence
are believed to contribute largely to the two-step process of the
$e^-e^+\rightarrow\overline NN$ reactions. The investigation of the reaction $e^-e^+\rightarrow\pi\omega$
reveals that the experimental data strongly prefer $\rho(1450)$ and $\rho(1700)$ being
dominantly $2S$ and $1D$ particles, respectively \cite{Kittimanapun}. One may correspondingly assign $\omega(1420)$
and $\omega(1650)$ to $2S$, $1D$ state or their mixtures. Therefore,
$\rho(2000)$ and $\omega(1930)$ are likely to be $3S$ or $2D$ states or mixtures of the
$3S$ or $2D$ states.

\begin{table}[h]
\begin{center}
\caption{Model parameters in the work.}
\vskip 0.5cm
\label{table2}
\begin{tabular}{|c|c|}
\hline
$a_\lambda(a_\rho)$ & 2.0 GeV$^{-1}$  \\
\hline
$b$ & 3.85 GeV$^{-1}$  \\
\hline
$\lambda_{25}$ & 1.0   \\
\hline
$\lambda_{34}$ & 2.6   \\
\hline
$\rho(2000)$ & $2000+i300$ GeV \\
\hline
$\omega(1930)$ & $1930+i150$ GeV \\
\hline
\end{tabular}
\end{center}
\end{table}

There are a number of model parameters which must be nailed down or taken from other works prior to any calculation which
may poss some predictions. Listed in Table \ref{table2} are the model parameters and their values employed in the work.
The size parameters $a_\lambda(a_\rho)$ of the radial wave function of nucleons are taken from the works \cite{Capstick,Chen}.
The size parameters $b$ of the radial wave function of the vector mesons $\rho^*$ and $\omega^*$ is let to be the same as the one
for the $\rho(770)$ meson, which is determined to be $b=3.85$ by the reaction $\rho(770)\rightarrow e^-e^+$.
Using as an input $M_\rho=0.7758$ GeV
and the experimental value of
$\Gamma_{\rho^0\rightarrow e^+e^-}=7.02\pm 0.11$ KeV, we get
$b=3.85$ GeV$^{-1}$
for the size parameter of the $\rho$ meson with the spatial wave
function set up in the harmonic oscillator
approximation \cite{yanNNbar,Kittimanapun}.
The size parameter $b$ might be slightly different from meson to meson.

The vertex $V_{25}$ is similar to the one for the reaction of one meson decay into
two mesons, thus we let $\lambda_{25}=1.0$ which gives the decay width of the
$\rho$ meson via the decay process $\rho\rightarrow\pi\pi$ in the $^3P_0$ models \cite{yanNNbar,Kittimanapun}.
However, the determination of the effective coupling constants $\lambda_{34}$ is
rather tentative. In this work we just let $\lambda_{34}=2.6$ which leads to
the branching decay width $\Gamma=31$ MeV for the reaction $\Sigma(1385)\rightarrow\Lambda\pi$
in the $^3P_0$ model where one uses $a=2.0$ GeV$^{-1}$ and $b=3.85$ GeV$^{-1}$ for
the involved baryons and meson, respectively. The masses and decay widths of the mesons $\rho(1930)$ and
$\omega(2000)$ are taken from the works \cite{rho2000,omega1930}.

\begin{figure}[h]
\centering
\includegraphics[height=0.4\textheight]{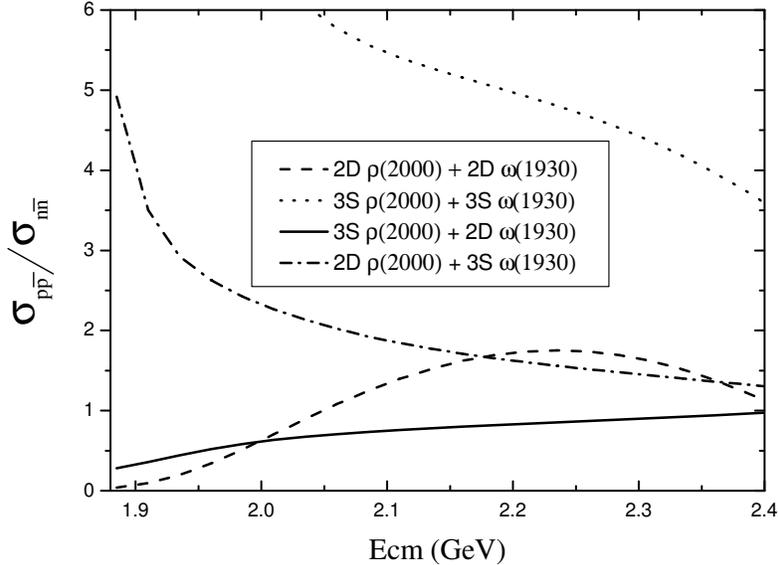}
\caption{Model predictions for the cross section ratio
$\sigma (e^+e^-\rightarrow\overline pp)/\sigma (e^+e^-\rightarrow\overline nn)$.
The solid, dashed, dash-dotted and dotted curves are for the results with
a $3S$ $\rho(2000)$ plus a $2D$ $\omega(1930)$, a $2D$ $\rho(2000)$ plus a $2D$ $\omega(1930)$,
a $2D$ $\rho(2000)$ plus a $3S$ $\omega(1930)$, and a $3S$ $\rho(2000)$ plus a $3S$ $\omega(1930)$,
respectively.\label{fig4}}
\end{figure}

In FIG.\ref{fig4} we give the model predictions for the ratio of the cross
section of the reaction $e^+e^-\rightarrow\overline pp$ to the one of the reaction $e^+e^-\rightarrow\overline nn$
in the two-step reaction. The ratio is independent of the effective coupling
constants $\lambda_{25}$ and $\lambda_{34}$ and insensitive to the nucleon and meson
size parameters $a$ and $b$.
Various combinations of $3S$ and $2D$ states for
the $\rho(2000)$ and $\omega(1930)$ mesons have been studied. It is found from
FIG. \ref{fig4} that
the ratio is well larger than 1 with a $S$-wave $\omega(1930)$ no matter whether
the $\rho(2000)$ meson is a $S$-wave or $D$-wave state. With $\omega(1930)$
being a $D$-wave meson, however, either a $S$-wave or $D$-wave $\rho(2000)$
gives a result, as shown in FIG. \ref{fig4}, in line with experimental data at $\overline NN$ threshold.
One may conclude that the experimental data of the cross section ratio rule out
a $S$-wave $\omega(1930)$ but have no
clear indication over the $\rho(2000)$ meson.

\begin{figure}
\centering
\includegraphics[height=0.25\textheight]{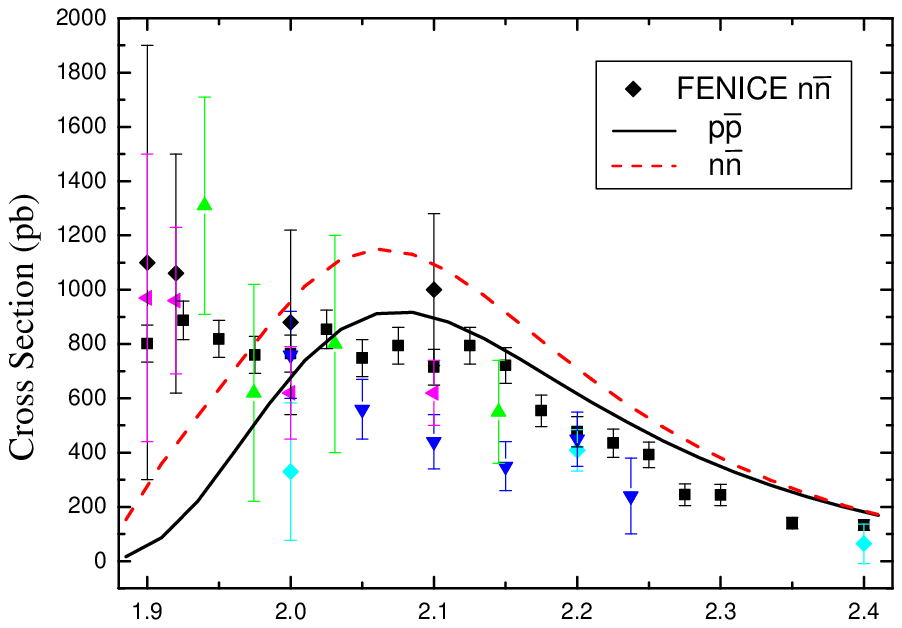}\\
\includegraphics[height=0.25\textheight]{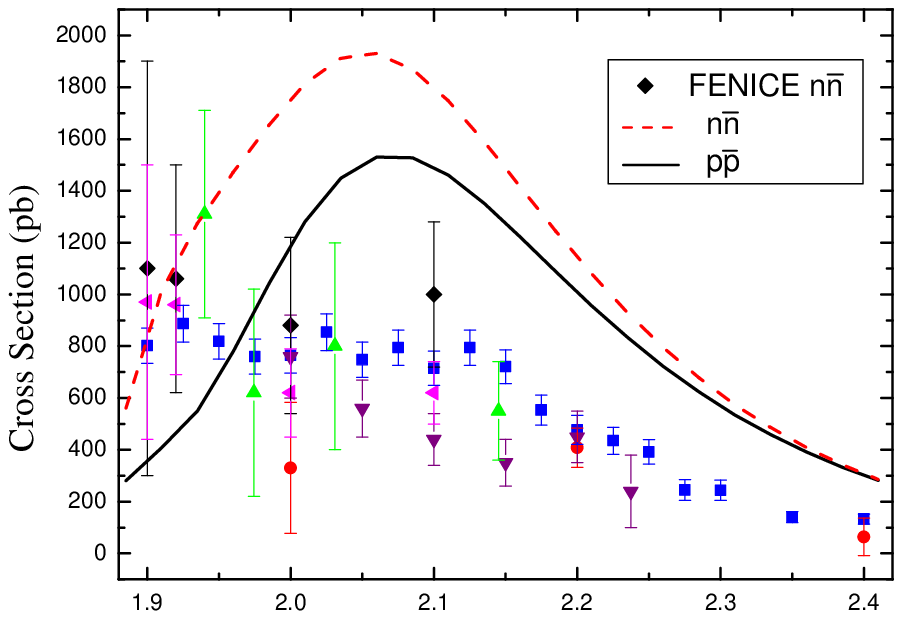}\\
\includegraphics[height=0.25\textheight]{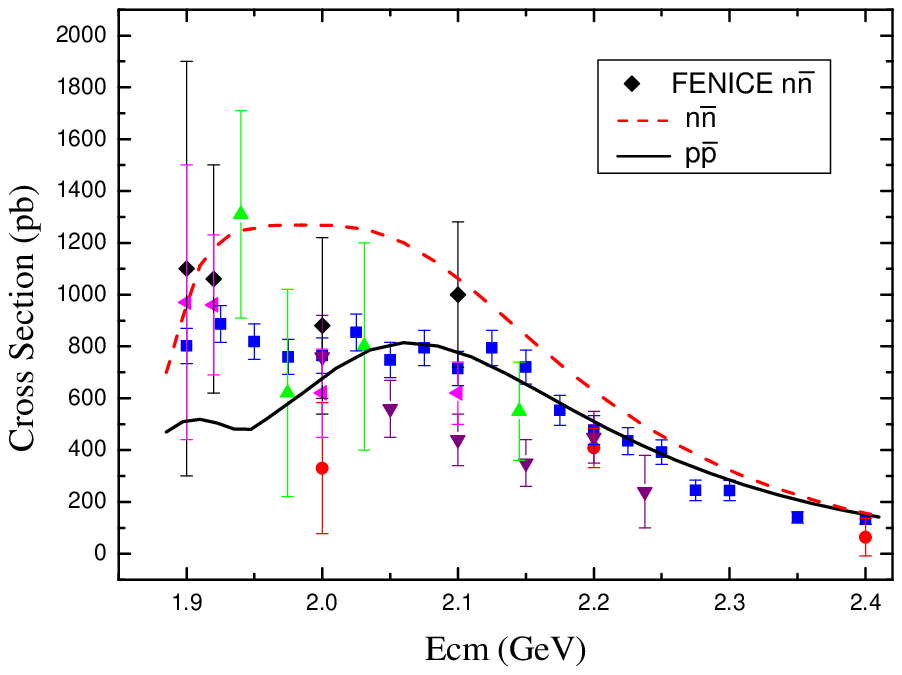}
\caption{Model predictions for the cross sections of the reactions
$e^+e^-\rightarrow\overline pp$ (solid curves) and $e^+e^-\rightarrow\overline nn$ (dashed curves):
Upper, middle and lower panels are for the two-step processes with a $3S$ $\rho(2000)$ plus a $2D$ $\omega(1930)$,
a $3S$ $\rho(2000)$ plus a $2D$ $\omega(1930)$ plus a $1D$ $\rho(1700)$, and a half $3S$ and half $2D$ $\rho(2000)$
plus a $2D$ $\omega(1930)$ plus a $1D$ $\rho(1700)$ as
the intermediate mesons, respectively. Solid diamonds are the FENICE experimental data on the reaction
$e^+e^-\rightarrow\overline nn$, while other symbols represent data on the reaction
$e^+e^-\rightarrow\overline pp$, as in FIG. \ref{fig1}.   \label{fig5}}
\end{figure}

The cross sections for the reactions $\sigma (e^+e^-\rightarrow\overline pp)$ and
$\sigma (e^+e^-\rightarrow\overline nn)$ are evaluated for two cases: (1) the intermediate
mesons $\rho(2000)$ and $\omega(1930)$ are respectively in the $3S$ and $2D$ states; (2)
the $\rho(2000)$ and $\omega(1930)$ mesons are both in the $2D$ state.
It is found that the theoretical results of the cross sections for both the reactions $e^+e^-\rightarrow\overline pp$ and
$e^+e^-\rightarrow\overline nn$ with a $2D$ $\rho(2000)$ and a $2D$ $\omega(1930)$
as the intermediate mesons is much smaller than the experimental data, and hence one may comfortably conclude that
the $\rho(2000)$ is unlikely to be a pure $D$-wave meson. Shown in the upper panel of FIG. \ref{fig5} are the
theoretical results for the cross section of the reactions $\sigma (e^+e^-\rightarrow\overline pp)$ and
$\sigma (e^+e^-\rightarrow\overline nn)$ with the intermediate mesons $\rho(2000)$ and $\omega(1930)$ being respectively
the $3S$-wave and $2D$-wave particles. The results are in line with the experimental data.

Considering that the $\rho(1700)$ is not far from the
$\overline NN$ threshold and posses a large decay width, it is necessary to have the $\rho(1700)$ meson
included as
an intermediate meson for the two-step process of the reactions $e^+e^-\rightarrow\overline NN$.
The investigation of the reaction $e^+e^-\rightarrow\omega\pi$ in the
$^3P_0$ model \cite{Kittimanapun} reveals that the $\rho(1700)$ meson is dominantly a $1D$-wave particle.
Shown in the middle panel of FIG. \ref{fig5} are the theoretical results with the intermediate mesons
$\rho(1700)$, $\rho(2000)$ and
$\omega(1930)$ being respectively in the $1D$, $3S$ and $2D$ states. In the calculation the mass
and width of the $\rho(1700)$ are taken from \cite{particletable} to be 1720 MeV and 250 MeV. Compared to the results
in the upper panel of FIG. \ref{fig5}, one finds that the $\rho(1700)$
gives a sizable contribution. The theoretical predictions with the intermediate mesons
$\rho(1700)$, $\rho(2000)$ and $\omega(1930)$
being respectively in $1D$, $3S$ and $2D$ states are obviously larger than the experimental data.

As mentioned above, the values for the effective coupling constants, particularly for $\lambda_{34}$, is
rather tentative. Therefore, one may consider to employ a little bit smaller coupling
constant $\lambda_{34}$ to lower down the theoretical results. However, it is found that though by employing
smaller coupling constants the theoretical results are improved for the region $E_{\rm c.m.} > 2.0$ GeV,
the behavior of the theoretical results close to the $\overline NN$ threshold is rather poor.
An alternative way to improve the theoretical predictions is to let the $\rho(2000)$ have some $D$-wave component.
Shown in the lower panel of FIG. \ref{fig5} are the theoretical results with the intermediate mesons
$\rho(1700)$ and $\omega(1930)$ being respectively pure $1D$ and $2D$ states but the $\rho(2000)$ being the mixture of
half $3S$ and half $2D$ waves. The theoretical results in the lower panel of FIG. \ref{fig5} are well in line with
the experimental data and, especially, the threshold behavior of the theoretical predictions are
largely improved.

\begin{figure}
\centering
\includegraphics[height=0.3\textheight]{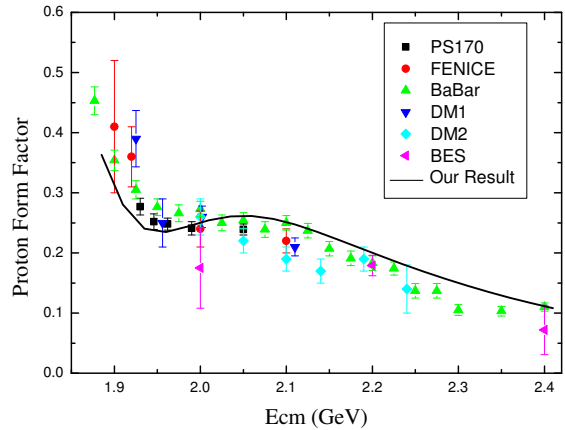}
\caption{Time-like proton form factor extracted from the
theoretical cross section for the reaction $e^+e^-\rightarrow\overline pp$ derived with
a $1D$ $\rho(1700)$, $2D$ $\omega(1930)$ and a half $3S$ and half $2D$ $\rho(2000)$ as
the intermediate mesons for the two-step process of the reactions $e^+e^-\rightarrow\overline NN$.\label{fig6}}
\end{figure}

The model prediction, as shown in FIG. \ref{fig6}, for the time-like proton form factor is extracted from the
theoretical cross section, shown in the lower panel of FIG. \ref{fig5},
of the reaction $e^+e^-\rightarrow\overline pp$, where
a $1D$ $\rho(1700)$, $2D$ $\omega(1930)$ and a half $3S$ and half $2D$ $\rho(2000)$ are employed as
the intermediate mesons for the two-step process of the reactions $e^+e^-\rightarrow\overline NN$.
The employed model
parameters, as listed in Table \ref{table2}, are all determined by other processes
or taken from other works. The theoretical prediction is well consistent with experimental data.

\section{Discussion and Conclusions}
The reactions $e^+e^-\rightarrow\overline NN$ at energy threshold
are studied in the
$^3P_0$ non-perturbative quark model.  The work suggests
that the two-step process, in which the primary $\overline qq$ pair
forms first a vector meson which in turn decays into a hadron pair,
is dominant over the one-step process in which the primary
$\overline qq$ pair is directly dressed by additional $\overline qq$
pairs to form a hadron pair.

The puzzling experimental result that
$\sigma(e^+e^-\rightarrow\overline
pp)/\sigma(e^+e^-\rightarrow\overline nn) < 1$ can be understood
in the framework of the phenomenological nonrelativistic quark
model. The theoretical
prediction for the time-like proton form factor at energy threshold
is well consistent with the experimental data.
All parameters employed in the model, except the ones
describing the mixture of the $S$ and $D$ waves for the
intermediate vector mesons $\rho(2000)$ and $\omega(1930)$, are
not free but determined by other reactions.

The experimental data of the reactions
$e^+e^-\rightarrow\overline nn$ and $\overline pp$ strongly suggest
the reported vector meson $\omega(1930)$ to be a $2D$-wave particle, while the
vector meson $\rho(2000)$ is preferred to be a mixture of the $3S$ and
$2D$ states.

\section*{Acknowledgments}
This work was supported in part by the Commission on Higher Education, Thailand (CHE-RES-RG Theoretical Physics).

\end{document}